\documentclass[12pt]{iopart}
\usepackage[latin1]{inputenc}
\usepackage[T1]{fontenc}
\usepackage[english]{babel}
\usepackage{latexsym}
\begin{document}
\title[Noether Symmetry of the Hyperextended Scalar Tensor theory for the FLRW models]{Noether Symmetry of the Hyperextended Scalar Tensor theory for the FLRW models}
\author{Stéphane Fay\\
14 rue de l'Etoile\\
75017 Paris\\
France\footnote[2]{Steph.Fay@Wanadoo.fr}}
\date{\today}
\begin{abstract}
We study in which conditions the Hyperextended Scalar Tensor theory in an FLRW background admits a Noether symmetry and derive the vectors field generating it.
\\
\\
\\
Published in Classical and Quantum Gravity copyright 2001 IOP Publishing Ltd\\
Classical and Quantum Gravity, Vol 18, 22, 2001.\\
http://www.iop.org
\end{abstract}
\maketitle
\section{Introduction} \label{s0}
One of the most well-known scalar tensor theories is the Brans-Dicke one \cite{BraDic61} developed in the sixties. This interest was motivated by the fact that it is able to reconcile a relativistic theory of gravitation with the ideas of Mach. It introduced the ideas of a gravitational coupling function, $G$, varying as the inverse of the scalar field $\phi$ and thus depending on the time\cite{Dir37} and of a coupling constant, $\omega$, between the scalar field and the metric functions. With the discovery of inflation by Guth \cite{Gut81} in the eighties allowing explaining why the Universe could be so flat or so isotropic, the interest for the scalar tensor theories has increased and found new justifications. Inflation could have been recently detected with the supernovae of type IA\cite{Per99, Rie98} and is often interpreted as the presence of a cosmological constant $\Lambda$ in the field equations. In the same time, physical particle progress have shown the importance of massive scalar field or varying gravitational functions. As instance the gravitational and the Brans-Dicke  coupling functions of the low energy effective string action are exponential laws of $\phi$. Moreover, unification theories predict a cosmological constant larger than the presently observed value. One way to solve this cosmological constant problem could be to consider a varying potential instead of a constant one. Thus the connection between particle physics and cosmology encourage us  to consider scalar-tensor theories more general than the Brans-Dicke one. The Lagrangian of the Hyperextended Scalar Tensor theories (HST) seems to be suited to take into account this need for generality since it is written with a gravitational function $G(\phi)$ and a Brans-Dicke coupling function $\omega(\phi)$. It is why we have chosen to study it when a potential $U(\phi)$ and a perfect fluid are presents.

The HST thus defined has then 3 undetermined functions. This is an advantage and a drawback at the same time. The advantage comes from the fact that any result we will obtain from this theory will be very general and could be applied to a large number of scalar tensor theories simply by assuming some special forms for $G$ and $\omega$. As instance, General Relativity with a scalar field is obtained with $G=G_0$ and Brans-Dicke theory for $G^{-1}=\phi$ and $\omega=\omega_0$. The drawback is that there are few indices indicating us what should be the physically interesting forms of the three undetermined functions depending on the scalar field. We can try to determine some of their characteristics from an observational point of view. Hence, in \cite{BoiEspPolSta00} it is shown how from the observations, it could be possible to determine the full Lagrangian and thus the potential from the luminosity distance and the linear density perturbation in the dust like matter as function of redshift. In \cite{SerAli96}, the convergence toward General Relativity, the presence of singularity or the dynamical evolution of the Universe at any time have been studied depending on the form of $\omega$. In \cite{IvaPotVar99}, observation of the variation of the fine-structure constant is analysed, giving us restriction on the possible variation of $G$. We can also leave the cosmological principle, assuming an anisotropic Universe and looking for the forms of the functions allowing the isotropy\cite{Fay01}. Considering the relations of this theory with the particle physics, another possibility is to claim that the HST could respect some of its symmetries as Noether symmetries. This is the approach chosen in \cite{ModKamBis99, RitMarPlaRubScuSto90} and that we will follow in this work. Our goal will be to look for the existence conditions of a Noether symmetry for the HST in different physical (in the vacuum (i.e. with a non-massive scalar field), with a potential or with a perfect fluid) and geometrical (flat open and closed Universe) contexts. Lets note that a transformation of the scalar field, $G^{-1}=\Phi$ allows to reduce the number of the undetermined function from three to two. The theory thus defined is named Generalised Scalar Tensor theory (GST) and has been studied from the same point of view in \cite{ModKamBis99, RitMarPlaRubScuSto90}. However, these results can not  be extended to HST by help of an inverse transformation and thus be related on important theories as the effective low energy string theory. The two classes of theories are physically equivalent but it is not possible to know the constraint imposed by the Noether symmetry on $G$, $U$ and $\omega$ from these determined for the GST.

Lets explain the interest of the Noether symmetries\footnote{Excellent tutorial from professors C. T. Hill and L. M. Lederman can be found on that subject in www.emmynoether.com.}. Noether symmetries theorem states that for every continuous symmetry of the laws of physics, there must exist a conservation law and reciprocally for every conservation law, there must exist a continuous symmetry. In this work, it will be studied via the approach of de Ritis et al \cite{RitMarPlaRubScuSto90} and Capozziello et al \cite{Cap94}. We will consider a point Lagrangian $L$ and a vector field $\chi$. A first step is to find the symmetry $\chi$, defined by the Lie derivative $\ell_\chi L$. The second step that we will not consider in this paper, is to determine the conserved quantity $Q$ that can be found by computing the Cartan one form $\theta _L=\frac{\partial L}{\partial \dot{a}}da+\frac{\partial L}{\partial \dot{\phi}}d\phi$, contracting it with $\chi$ to get $Q=i_\chi\theta_L$. Then, the calculus of $Q$ can allow to solve exactly the field equations as shown in the previously quoted papers. Thus Noether symmetry is very important in the search for exact solutions of theories having particular symmetries and conserved quantities, helping the study of more general ones. However, note that recently, it has been demonstrated that Noether symmetry could, for some cases, not be consistent with dynamical equations and that other type of symmetries could exist\cite{SanMod01} indicating that in the future we will have to proceed with care when we consider the second step.\\

The geometrical framework of this study will be the FLRW models. It would be more logical to consider an anisotropic and inhomogeneous model more qualified to describe the geometry of the early Universe where particle physics naturally takes place. However, it does not exist a full classification of these geometries contrary to the FLRW or Bianchi models. The relative simplicity of the FLRW models will allow us to study the Noether symmetries for a whole class of geometrical models and thus we will be able to make comparisons between each of them.

The plane of this paper is the following: in the section \ref{s1}, we look for the conditions allowing a Noether symmetry for the HST in the FLRW models. We discuss our results and conclude in the section \ref{s2}.
\section{Noether symmetry of the FLRW} \label{s1}
We use the following form of the metric describing an isotropic and homogeneous Universe:
\begin{equation} \label{0}
ds^2=-dt^2+a^2d\Omega
\end{equation}
$a$ being the scale factor. The Lagrangian of the HST with a potential and a perfect fluid is written: 
\begin{equation} \label{1}
L=\left[G(\phi)^{-1}R+\omega\phi^{-1}\phi_{,\mu}\phi^{,\mu}-U\right]\sqrt{-g}+L_m
\end{equation}
$G$ being the gravitational coupling function, $\omega$ the Brans-Dicke coupling function, $U$ the potential, $\phi$ the scalar field from which depends on previous quantities and $L_m$ the Lagrangian  corresponding to a perfect fluid with equation of state $p=(\delta-1)\rho$. Using the fact that $\int\Box G^{-1}\sqrt{-g}=0$, the point Lagrangian for the FLRW models is written: 
\begin{equation} \label{3}
L=-6 G^{-1} a\dot{a}^2-6G^{-1}_\phi a^2\dot{a}\dot{\phi}+\omega a^3\phi^{-1}\dot{\phi}^2+6ka G^{-1}-a^3 U+\rho_0(\gamma-1)a^{3(1-\gamma)}
\end{equation}
To find the conditions for Noether symmetry, we will follow the approach of de Ritis et al \cite{RitMarPlaRubScuSto90} and Capozziello et al \cite{Cap94}. We will consider the configuration space $E=(a,\phi)$ whose corresponding tangent space is $TE=(a,\dot{a},\phi,\dot{\phi})$. The vector field $X$ generating the symmetry is then: 
\begin{equation}
X=\alpha\frac{\partial }{\partial a}+\chi \frac{\partial }{\partial \phi}+\dot{\alpha} \frac{\partial }{\partial \dot{a}}+\dot{\chi}\frac{\partial }{\partial \dot{\phi}}
\end{equation}
where $\alpha$ and $\chi$ are some functions of $a$ and $\phi$. The existence of a Noether symmetry induces the existence of the vectors field $X$  such that $\ell _X L=0$, $\ell_X$ being the Lie derivative with respect to $X$. The meaning of this equation is that $L$ is constant along the flow generated by $X$. We deduce from it a second-degree expression for $\dot{a}$ and $\dot{\phi}$ whose coefficients only depend on $a$ and $\phi$ and have to vanish. 
\subsection{Vacuum model} \label{s11}
Applying the principle described above, we get the following equations when no potential or perfect fluid is present: 
\begin{eqnarray}
&k(G^{-1}\alpha+(G^{-1})_\phi a\chi)=0 \label{3a}&\\
&G^{-1}\alpha+(G^{-1})_{\phi}a\chi+2G^{-1}a\frac{\partial\alpha}{\partial a}+(G^{-1})_\phi a^2\frac{\partial \chi}{\partial a}=0\label{3b}&\\
&3\omega\alpha-\omega\phi^{-1} a\chi-6\phi(G^{-1})_\phi\frac{\partial \alpha}{\partial \phi}+2\omega a\frac{\partial \chi}{\partial \phi}+\omega_\phi a\chi=0\label{3c}&\\
&6(G^{-1})_\phi\alpha+3(G^{-1})_{\phi\phi}a\chi+3(G^{-1})_\phi a\frac{\partial \alpha}{\partial a}+6G^{-1}\frac{\partial \alpha}{\partial \phi}-\omega\phi^{-1} a^2\frac{\partial \chi}{\partial a}+3(G^{-1})_\phi a\frac{\partial \chi}{\partial \phi}=0\label{3d}&\\\nonumber
\end{eqnarray}
They are the same as these of \cite{ModKamBis99} when we choose $G^{-1}=\phi$. We are going to examine successively the case of curved and flat models. 
\subsubsection{$k\not =0$} \label{s111}
Following the methods of \cite{ModKamBis99}, from the previous equations system, it should be possible to determine a relation between $\omega$ and $G$ necessary for the existence of a symmetry and the forms of $\alpha$ and $\chi$ determining the vector field generating it. Starting from (\ref{3a}), we deduce that $\alpha=-G(G^{-1})_\phi a\chi$. Then, putting this quantity in (\ref{3b}) and integrating, we get:
\begin{eqnarray} \label{3f2}
\chi&=&n(\phi)a^{-2}\\
\alpha&=&-n(\phi)G(G^{-1})_\phi a^{-1}\\\nonumber
\end{eqnarray}
with $n(\phi)$ a function of the scalar field that we have to determine. We introduce these expressions in (\ref{3c}) and we get the relation we are looking for between $\omega$ and $G$:
\begin{equation} \label{3f}
\omega=3\omega_0\phi (G^{-1})_\phi^2G(G^{-2}-2\omega_0)^{-1}
\end{equation}
$\omega_0$ being an integration constant. To obtain $n(\phi)$, we replace $\omega$ in (\ref{3d}) by this last expression. We find that:
\begin{equation} \label{3f1}
n(\phi)=n_0(2\omega_0-G^{-2})(G^{-1})_\phi^{-1}
\end{equation}
$n_0$ being an integration constant. We conclude that in the vacuum case and for a curved FLRW model, a HST whose Brans-Dicke coupling function is linked to the gravitational function by the relation (\ref{3f}) has a Noether symmetry generated by a vectors field $X$ defined by (\ref{3f1}) and (\ref{3f2}). These relations generalise these found in the case of GST in \cite{ModKamBis99}.\\
\subsubsection{$k =0$} \label{s112}
The equation (\ref{3a}) vanishes and then we have 4 undetermined quantities (the partial derivatives of $\alpha$ and $\chi$) and three equations. To find some solutions, we assume that $\alpha$ and $\chi$ can be written with separating variables as successfully done in \cite{ModKamBis99} in the case of GST and we define $\alpha=\alpha_1(a)\alpha_2(\phi)$ and $\chi=\chi_1(a)\chi_2(\phi)$. Introducing these expressions in (\ref{3b}), we find that it will be satisfied if $\alpha_2=mG(G^{-1})_\phi\chi_2$, $m$ being a constant. Then we do the same thing with (\ref{3c}) and see that we must have $\chi_1=na^{-1}\alpha_1$. Introducing these expressions for $\alpha_2$ and $\chi_1$ in the equations (\ref{3b}-\ref{3d}), we deduce from the two first ones the expressions for $\alpha_1^{-1}d\alpha_1/da$ and $\chi_2^{-1}d\chi_2/d\phi$ that we use in the last equations. We get the following relation that have to be satisfied between $\omega$ and $G$:
\begin{equation} \label{3i}
\omega=\phi G(G^{-1}_\phi)^2\frac{6m\omega_0(m+n)G^{3m/n+1}-3}{n^2\omega_0G^{3m/n+1}+2}
\end{equation}
So that $\alpha$ and $\chi$ be determined we calculate that:
\begin{equation}
\alpha_1=\alpha_{10}a^{-m(2m+n)^{-1}}
\end{equation}
\begin{equation}
\chi_2=\chi_{20}(G^{-1})_\phi^{-1}G^{\frac{m(3m+2n)}{n(2m+n)}}(n^2\omega_0+2G^{-3m/n-1})^{1/2}
\end{equation}
$\alpha_{10}$ and $\chi_{20}$ being integration constants. Consequently a HST whose gravitational function and Brans-Dicke coupling functions are linked by the relation (\ref{3i}) have a Noether symmetry generated by the vectors field defined by $\alpha_1$, $\alpha_2$, $\chi_1$ and $\chi_2$.\\
\subsection{HST with potential} \label{s12}
When we consider a potential, only the first equation of (\ref{3a}-\ref{3d}) changes and is written:
\begin{equation}\label{3h}
6kG^{-1}\alpha- 3 a^2 U \alpha+ 6 k a \chi (G^{-1})_\phi - a^3 \chi U_\phi=0
\end{equation}
Let's consider a model with and without curvature.
\subsubsection{$k\not = 0$} \label{s121}
Using the same method as in section \ref{s11}, we express $\alpha$ with (\ref{3h}) and put its expression in (\ref{3b}) to determine this of $\chi$. It comes:
\begin{equation} \label{3k}
\chi=n(\phi )(2kG^{-1} -a^2 U) \mbox{[}3a^2U(G^{-1})_\phi +  G^{-1}( 6k(G^{-1})_\phi - 2a^2U_\phi) \mbox{]}^{\frac{-3U(G^{-1})_\phi+3U_\phi G^{-1}}{6U(G^{-1})_\phi-4U_\phi G^{-1}}}a^{-2}
\end{equation}
$n(\phi)$ being a function depending on the scalar field. If we Introduce the forms of  $\alpha$ and $\chi$ in (\ref{3c}) we get a differential equation for $n(\phi)$ which is written with 3 different powers of $a$ and logarithm of an expression of $a$ and $\phi$. This equation having to vanish, the coefficient of the logarithmic expression has to be equal to zero. This is only possible if the power of the expression for $\chi$ is a constant $F_0$, i.e. when $U=U_0G^{-\frac{3(1+2F_0)}{3+4F_0}}$, $U_0$ being an integration constant. In the same way the coefficient of the powers of $a$ have to be equal to zero thus defining a system of 3 equations whose the only solution is the General Relativity with $G=G_0$ and $\omega=0$. However, for this theory $\chi$ is undetermined and thus we conclude that for a massive HST in a curved Universe, there is no Noether symmetry. As shown in \cite{SanMod01}, it does not mean that there is no symmetry at all, and then conserved quantities, but only for the special one we have considered and which belongs to the class of point symmetries\cite{RitMarPlaRubScuSto90}.
\subsubsection{$k= 0$} \label{s122}
Contrary to what happens for the vacuum case, the equation (\ref{3a}) is not identically zero but is written:
\begin{equation} \label{3j}
3 U \alpha+ a \chi U_\phi=0
\end{equation}
It follows that we have not to assume a variable separation for $\alpha$ and $\chi$. From this last equation, we deduce: 
\begin{equation} \label{3x1}
\alpha=1/3a\chi U^{-1} U_\phi
\end{equation}
We introduce this result in (\ref{3b}) and derive $\chi$:
\begin{equation} \label{3x2}
\chi=n(\phi)a^{\frac{3(U(G^{-1})_\phi-G^{-1}U_\phi)}{-3U(G^{-1})_\phi+2G^{-1}U_\phi}}
\end{equation}
$n(\phi)$ being a function of the scalar field. In the equation (\ref{3d}), we replace $\alpha$ and $\chi$ by their forms above  thus getting a differential equations for $n(\phi)$. Its form is $F_1(\phi)n_\phi+F_2(\phi)+F_3(\phi)ln(a)=0$. To satisfy it, it is necessary that $F_3=0$ thus implying $U=U_0G^{-p}$ with $p$ a constant or $G=G_0$, which corresponds to General Relativity with a massive scalar field. This two cases allow independently that $F_3=0$ and are independents each others since in the second one, their is no constraint between $G$ and $U$. We examine successively these two cases.\\
\underline{When $U=U_0G^{-p}$}, we get an expression for $n_\phi$ from (\ref{3d}). Introducing $\alpha$, $\chi$ and $n_\phi$ in the equation (\ref{3c}), we get the following relation between $G$ and $\omega$:
\begin{equation} \label{24}
\omega=\frac{\phi \left[ -3 + 2p(3-p) \omega_0 G^{-p+1}\right](G^{-1})_\phi^2}{2G^{-1} - 3\omega_0G^{-p}}
\end{equation}
$\omega_0$ being an integration constant. Using this last expression, we derive the exact form of $n$:
\begin{equation} \label{3x3}
n=n_0G^{\frac{p(-6p^2+p+3)}{2(p-1)(3+2p)^2}}(2G^{-1}-3\omega_0G^{-p})^{\frac{(1+p)(3-2p)^2}{2(p-1)(3+2p)^2}}
\end{equation}
$n_0$ being an integration constant. Consequently a massive HST whose potential is proportional to a power of the gravitation function which itself depends on the Brans-Dicke coupling function by the relation (\ref{24}) have a Noether symmetry generated by a vectors field $X$ determined by the relations (\ref{3x1}-\ref{3x3}).\\
If we consider the General Relativity, a Noether symmetry only exists in presence of a cosmological constant. However $\chi$ is not determined and thus there is no symmetry.\\
\underline{When $G=G_0$}, we calculate from the equations (\ref{3c}) and (\ref{3d}) that a Noether symmetry will exist if:
\begin{equation} \label{20}
\omega=\frac{2\phi(U_\phi)^2}{G_0(\omega_0+3U)U}
\end{equation}
Then, from (\ref{20}), we derive the form of $n$:
\begin{equation} \label{3x4}
n=n_0U(\omega_0+3U)^{1/2}(U_\phi)^{-1}
\end{equation}
Thus a Noether symmetry can exist for General Relativity with a massive scalar field different of a constant in a flat Universe if the relation (\ref{20}) between the potential and the Brans-Dicke coupling function is satisfied and is generated by a vector fields $X$ defined by (\ref{3x1}), (\ref{3x2}) et (\ref{3x4}).\\
\subsection{HST with a perfect fluid} \label{s13}
The term representing a perfect fluid with an equation of state $p=(\gamma-1)\rho$ in the Lagrangian is $\rho_0(\gamma-1)a^{3(1-\gamma)}$. Once again, only the equation (\ref{3a}) is modified and written:
\begin{equation}\label{28}
2k G^{-1}\alpha+ 2ak\chi (G^{-1})_\phi-\rho_0 (\gamma-1)^2 a^{2-3\gamma}\alpha
\end{equation}
We deduce from it an expression for $\alpha$:
\begin{equation}\label{28x1}
\alpha= -2ak\chi (G^{-1})_\phi(2k G^{-1}-\rho_0 (\gamma-1)^2 a^{2-3\gamma})^{-1}
\end{equation}
that we introduce in (\ref{3b}). Then, we get for $\chi$:
\begin{equation} \label{28x2}
\chi=n(\phi)a^{\frac{4-3\gamma}{-2+3\gamma}}\left[(\gamma-1)^2\rho_0a^2-2ka^{3\gamma}G^{-1}\right]\left[(\gamma-1)^2\rho_0a^2+2ka^{3\gamma}G^{-1}\right]^{\frac{1-3\gamma}{-2+3\gamma}}
\end{equation}
$n(\phi)$ being a function depending on the scalar field. We calculate it by using the expressions for $\alpha$ and $\chi$ in (\ref{3d}). To this end, we consider three values of $\gamma$, 1, 0 and 4/3 corresponding respectively to a dust dominated, vacuum energy dominated and radiation dominated Universe. In the first case, the equation (\ref{28}) takes the same form as in the vacuum and thus the results are the same as these of section \ref{s11}. In what follows, we will consider only the two last ones.
\subsubsection{$k\not =0$} \label{s131}
When $\gamma=0$ or $\gamma=4/3$, if we introduce the expressions for $\alpha$ and $\chi$ in the equation (\ref{3d}), we get a polynomial expression for $a$ whose coefficients have to be zero. It corresponds to a system of 3 equations with 3 unknowns $G$, $\omega$ and $n$ that we have to determined.\\
When the Universe is vacuum dominated, the only possible solution corresponds to General Relativity or $n=0$ but once again $\chi$ is undetermined and a Noether symmetry does not exist.\\
When the Universe is radiation dominated, the equations are satisfied if:
\begin{equation} \label{28x3}
n=n_0\sqrt{G^{-1}}(G^{-1})_\phi^{-1}
\end{equation}
\begin{equation} \label{28x4}
\omega=-3/2\omega_0\phi G(G^{-1})_\phi^2
\end{equation}
Thus, the HST with a perfect radiative fluid can have a Noether symmetry if this relation between the gravitational coupling function and the Brans-Dicke coupling function is satisfied. It is generated by the vectors field $X$ defined by (\ref{28x1}-\ref{28x3}).\\
\subsubsection{$k=0$} \label{s132}
For a flat Universe, the equation (\ref{28}) shows that $\alpha=0$. Then we can calculate that $\chi=n(\phi)a^{-1}$ whatever $\gamma\not =1$ with:
\begin{equation} \label{28x5}
n(\phi)=n_0\sqrt{\omega_0-2G^{-1}}(G^{-1})_\phi 
\end{equation}
\begin{equation} \label{28x6}
\omega=3\phi (G^{-1})_\phi^2(\omega_0-2G^{-1})^{-1}
\end{equation}
It follows that for a flat model with a perfect fluid, the HST admit a Noether symmetry when the equation (\ref{28x6}) is satisfied. It is then generated by the vectors field $X$ defined by (\ref{28x1}), (\ref{28x2}) and (\ref{28x5}).
\section{Discussion} \label{s2}
In this work, we have studied the Noether symmetries of the HST for the FLRW models in vacuum, with a massive potential or with a perfect fluid. Our results consist in the determination of conditions allowing the existence of symmetry. They are relations between the functions $G$, $\omega$ or $U$, conditions on their forms or on the type of geometry. Moreover, for each case, we have determined the vectors field $X$ generating the symmetry.\\

When we consider a HST without any potential or perfect fluid, for a curved or flat geometry, a Noether symmetry will exist if respectively the relations (\ref{3f}) or (\ref{3i}) are respected between the gravitational and Brans-Dicke coupling functions. They generalise these found in \cite{RitMarPlaRubScuSto90}.\\
When we consider a potential, a symmetry can only exist for a flat Universe. The potential have to be proportional to a power of the gravitational coupling function or the theory to correspond to the General Relativity with a massive scalar field different from a cosmological constant. Respectively, a relation between $\omega$ and $G$ defined by (\ref{24}) or $\omega$ and $U$ defined by (\ref{20}) have to be satisfied.\\
When we consider a perfect fluid, for a dust dominated Universe the results are the same as these of the vacuum case. For a curved Universe, if the Universe is vacuum energy dominated, the symmetry does not exist. If it is radiation dominated, a relation between $\omega$ and $G$ defined by (\ref{28x4}) have to be respected. For the existence of a Noether symmetry for a flat Universe and for any type of matter, the relation that is imposed by the symmetry is given by (\ref{28x6}). These results are summarised in the table 1. For each of these cases, we have calculated all the elements allowing the determination of the vectors field $X$ generating the Noether symmetry.\\

Lets discuss about interesting theories. Whatever the relation existing between $G$, $\omega$ and $U$ for the symmetry existence, the General Relativity defined by $G=G_0$ and $\omega=0$ with or without a cosmological constant always respects it. However, in this case, the quantity $\chi$ is never defined and the General Relativity has no symmetry as conclude in \cite{Kuc81}. The only case for which General Relativity admits a Noether symmetry is in presence of a massive scalar field, the potential being different from a cosmological constant, in a flat Universe. Therefore, we recover and generalise the result of \cite{RitMarPlaRubScuSto90} which corresponds to the theory defined by $\omega=\phi$, and then to an exponential potential.\\
If we consider a GST defined by $G^{-1}=\phi$, in the vacuum case, the only GST admitting a Noether theory is defined by $\omega=3m\omega_0(\phi^2-2\omega_0)^{-1}$ for a curved Universe and $\omega=\left[-3+6m\omega_0(m+n)\phi^{-3m/n-1}\right](2+n^2\omega_0\phi^{-3m/n-1})^{-1}$ for a flat Universe as it has been shown in \cite{ModKamBis99} where the dynamics of these two theories are studied. If we consider a potential in a flat Universe, we note that its only form allowing a symmetry is a power law of the scalar field. Then, the Brans-Dicke coupling function have to be $\omega=\left[-3+2p(3-p)\omega_0\phi^{p-1}\right](2-3\omega_0\phi^{p-1})^{-1}$. It is interesting to note that it is the same form as this of the GST in the empty. If we consider the presence of a radiative fluid, the Brans-Dicke theory is the only GST allowing a symmetry for a Universe with a curvature.\\ If we consider a theory whose gravitational function is defined by $e^{-\phi}$ and corresponding to the form usually used for the effective string theory action at low energy, we remark that for a flat Universe the only type of potential allowing a symmetry will be an exponential law of the scalar field. The Brans-Dicke coupling function is then $\omega=\phi e^{-\phi}\left[3+2p\omega_0(p-3)e^{(p-1)\phi}\right](-2+3\omega_0 e^{(p-1)\phi})^{-1}$ and is quite different from the $\omega$ usually used for this type of theory although it asymptotically tends toward it.\\

In a general way, we note that the type of geometry favouring a Noether symmetry is rather a flat one. To our knowledge, such an attempt to draw up a complete list of the HST, in various physical and geometrical contexts, admitting a Noether symmetry, does not exist in the literature. A next step will be to look for conserved quantities and then to try to determine the main dynamical characteristics of classes of models thus defined as it has been done for the special cases of GST in \cite{ModKamBis99}. Here, since our goal was to find the conditions for the existence of Noether symmetries, we have not undertaken this task that will be probably too large for a single article. Another possible extension will be to consider Bianchi models.
 
\begin{table}[p]
\caption{Classification of the Hyperextended Scalar Tensor theories admitting a Noether symmetry.}
\begin{center}
\begin{tabular}{lll}
\br
&$k\not =0$&$k=0$\\
\mr
Empty&$\omega=3\omega_0\phi (G^{-1})_\phi^2G(G^{-2}-2\omega_0)^{-1}$&$\omega=\left[-3+6m\omega_0\phi^{-3m/n-1}(m+n)\right]$\\
&&$(2+n^2\omega_0\phi^{-3m/n-1})^{-1}$\\
\mr
Potential&No symmetry& If $U=U_0G^{-p}$, $\omega=\frac{\phi \left[ -3 + (6-2p) p \omega_0 G^{-p+1}\right](G^{-1})_\phi^2}{2G^{-1} - 3\omega_0G^{-p}}$\\
&&If $G=G_0$, $\omega=\frac{2\phi(U_\phi)^2}{G_0(\omega_0+3U)U}$\\
\mr
Dust: $\gamma=1$&Same as for empty&Same as for empty\\
Vacuum: $\gamma=0$&No symmetry&$\omega=3\phi (G^{-1})_\phi^2(\omega_0-2G^{-1})^{-1}$\\
Radiation: $\gamma=4/3$&$\omega=-3/2\omega_0\phi G(G^{-1})_\phi^2$&$\omega=3\phi (G^{-1})_\phi^2(\omega_0-2G^{-1})^{-1}$\\
\br
\end{tabular}
\end{center}
\end{table}

\end{document}